\documentclass[aps,prb,twocolumn,superscriptaddress,showpacs]{revtex4-1}

\usepackage{amsmath,amssymb,graphics,epsfig,epstopdf,color,verbatim,ulem,braket,tabularx}
\usepackage[colorlinks,linkcolor=blue,citecolor=blue,urlcolor=blue]{hyperref}

\begin{document}

\title{Correlated Dirac semimetal by periodized cluster dynamical mean-field theory}

\author{Qing-Xiao Li}
\affiliation{Department of Physics, Renmin University of China, Beijing 100872, China}
\author{Rong-Qiang He}\email{rqhe@ruc.edu.cn}
\affiliation{Institute for Advanced Study, Tsinghua University, Beijing 100084, China}
\affiliation{Department of Physics, Renmin University of China, Beijing 100872, China}
\author{Zhong-Yi Lu}\email{zlu@ruc.edu.cn}
\affiliation{Department of Physics, Renmin University of China, Beijing 100872, China}

\begin{abstract}
  The periodized cluster dynamical mean-field theory (PCDMFT) combined with exact diagonalization as impurity solver has been applied to the half-filled standard Hubbard model on the honeycomb lattice. A correlated Dirac semimetal is found for weak interactions and it transforms into an antiferromagnetic insulating phase for strong interactions via a first-order quantum phase transition, not intervened by a spin liquid phase in between. In this application, the PCDMFT introduces the partial translation symmetry, but cures well the problem due to the translation symmetry breaking in the cluster dynamical mean-field theory studies for the same model, which give rise to a spurious insulating phase in the weakly interacting region.
\end{abstract}


\date{\today}
\maketitle

\section{Introduction}

In past decades, much effort has been devoted to searching for materials and models that show a quantum spin liquid, in which the local moments fluctuate even at zero temperature. Such a state is a genuine Mott insulator without spontaneously breaking spatial and spin symmetries and can show remarkable emergent phenomena such as non-trivial topological order, fractionalized charges and excitations. A common idea is that geometric frustration may prevent orders and lead to spin liquids~\cite{Balents10}, such as in models on kagome lattices. Another proposal is that a spin liquid may emerge nearby a Mott transition~\cite{Lee05,Hermele07}, such as in the Hubbard models on triangular or honeycomb lattices. Meng and coworkers ~\cite{Meng10} attempted the later idea. They performed large-scale quantum Monte Carlo (QMC) calculations on a standard  Hubbard model at half-filling on the honeycomb lattice at zero temperature. After the obtained QMC results for finite-size clusters containing up to $18 \times 18 \times 2$ sites were carefully extrapolated to the thermodynamic limit, they found a spin liquid state between the Dirac semimetal at weak interactions and the antiferromagnetic insulator at strong interactions. But, later this result was challenged by Sorella et al.~\cite{Sorella12} with the similar QMC calculations on larger clusters with up to $36 \times 36 \times 2$ sites and by Assaad and Herbut~\cite{Assaad13} with an advanced QMC technique.

On the other hand, the existence of spin liquid on the honeycomb lattice has also been studied~\cite{Wu10,Liebsch11,Yu11,He12,Seki12,Hassan13} by using quantum cluster methods~\cite{Maier05}, which map a lattice model, a Hubbard model for example, onto a cluster of impurity sites coupled to a set of bath sites determined self-consistently or by a variational principle. Wu et al.~\cite{Wu10} used cluster dynamical mean-field theory~\cite{Kotliar01} (CDMFT) combined with continuous-time quantum Monte Carlo~\cite{Rubtsov05} (CTQMC) as impurity solver, while Liebsch~\cite{Liebsch11} also used CDMFT but with exact diagonalization~\cite{Caffarel94} (ED) as impurity solver. They both did calculations at finite temperatures and showed the existence of a spin liquid and a Mott gap at intermediate interactions. However, performing also ED-CDMFT but at zero temperature and with elaborate numerical analytic continuation on the Matsubara Green's functions, He and Lu~\cite{He12} found that the existence of the single-particle spectral gap extends actually to the weak interaction limit, i.e., for all interactions $U > 0$. Later, this result was reproduced by Seki and Ohta~\cite{Seki12}, who used variational cluster approximation~\cite{Potthoff03a,Potthoff03b} (VCA) with also ED as impurity solver. Since it is known that the Dirac semimetal is stable for weak interactions~\cite{Giuliani09,Honerkamp08}, the validity of the application of CDMFT-like methods to the present model is questioned~\cite{Hassan13,Liebsch13,Liebsch13comment,Hassan13reply}.

In all works using quantum cluster methods mentioned above, the impurity clusters consist of six sites forming a ring with the same rotation symmetries as the honeycomb lattice. When the ED is used as the impurity solver, only six bath sites couple to the impurity cluster, which is close to the ED's computational limit. Applying VCA, CDMFT, and the cluster dynamical impurity approximation~\cite{Potthoff12book} (CDIA) at zero temperature with the ED as the impurity solver to honeycomb and square lattices, Hassan and S\'en\'echal~\cite{Hassan13} argued that one bath site per boundary impurity site in ED CDMFT and ED VCA is inadequate and leads to the absence of the semimetallic phase for the honeycomb lattice. In contrast, when they adopted two- or four-site impurity clusters with two bath sites per boundary impurity site, a first-order transition at a finite interaction from the semimetal to the antiferromagnetic insulator was obtained. However, this argument was refuted by Liebsch et al.~\cite{Liebsch13comment,Liebsch13}, who showed that CTQMC CDMFT, which uses a continuum of bath sites, yields the same results as the ED CDMFT at finite but low temperatures for the honeycomb lattice. Moreover, with the detailed analysis~\cite{Liebsch13comment,Liebsch13} on the ED CDMFT calculations, they attributed the absence of the semimetallic phase to the translation symmetry breaking in the CDMFT.

In order to avoid the translation symmetry breaking, Liebsch and Wu~\cite{Liebsch13} resorted to dynamic cluster approximation~\cite{Hettler98,Hettler00} (DCA), which is a quantum cluster method preserving translation symmetry naturally. After using the DCA, the semimetallic phase at weak interactions was successfully recovered. However, they found that this semimetallic phase is still robust near $U = 6t$ at which the CDMFT and large-scale QMC calculations already give rise to a Mott insulating phase. This shows that the DCA overemphasizes the semimetallic behavior, which is mostly because its condition on ensuring translation symmetry is too rigid for the description of correlations within the impurity cluster. Thus neither DCA nor CDMFT can describe the correlations properly in the whole interaction range on the honeycomb lattice.

Here we revisit the problem, namely, the quantum phases of the half-filled standard Hubbard model on the honeycomb lattice, with periodized CDMFT (PCDMFT)~\cite{Capone04,Biroli04}, to realize a proper description of the semimetallic phase at weak interactions and meanwhile retain CDMFT's advantage in well describing correlations at strong interactions in the framework of quantum cluster method. The PCDMFT is a variation of CDMFT partially restoring translation symmetry. As shown below, the PCDMFT indeed well reproduces the semimetallic phase at weak interactions, separated from the antiferromagnetic insulating phase at strong interactions by a first-order phase transition with the $U_c \sim 3.7t$, and there isn't a spin liquid emerging near the transition. These are in agreement with the results from the large-scale QMC calculations by Sorella et al.~\cite{Sorella12}

\section{Model and method}
\begin{figure}[h]
  \includegraphics[width=\columnwidth]{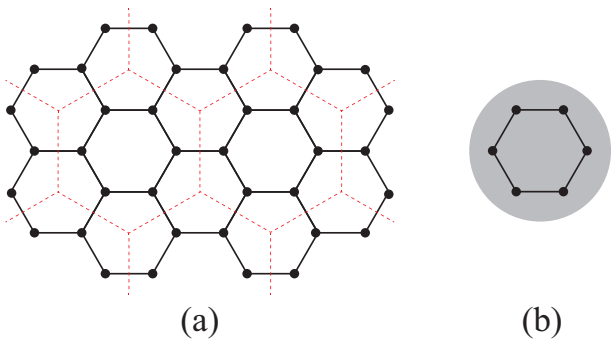}
  \caption{(Color online) (a) The honeycomb lattice is a lattice with each primitive unitcell consisting of two sites. The CDMFT/PCDMFT considers it as a superlattice with six-site rings as supercells, for example. (b) The CDMFT/PCDMFT maps the Hubbard model on the honeycomb lattice onto an impurity model by considering a supercell as an impurity cluster coupling to a non-interacting bath (schematically shown by the shaded area) which approximates all other supercells.}
  \label{fig:weiss}
\end{figure}

As in those works discussed above, we study the half-filled Hubbard model on the honeycomb lattice as follows,
\begin{equation}\label{eq:hamiltonian}
\hat{H} = -t\sum_{\langle ij\rangle,\sigma}(c_{i\sigma}^{\dagger}c_{j\sigma}+c_{j\sigma}^{\dagger}c_{i\sigma}) + U\sum_{i}n_{i\uparrow}n_{i\downarrow},
\end{equation}
where $\sigma = \uparrow$ or $\downarrow$, $c_{i\sigma}$ annihilates an electron with spin $\sigma$ at the $i$th lattice site, $n_{i\sigma}$ is the number of electrons with spin $\sigma$ at site $i$, and $\langle ij\rangle$ denotes a pair of nearest-neighbor sites $i$ and $j$. The studied honeycomb lattice is shown in Fig.~\ref{fig:weiss}.

The CDMFT considers the lattice as a superlattice composed of supercells, each of which consists of a cluster of sites, as shown in Fig.~\ref{fig:weiss}. Then the CDMFT takes a cluster as the reference cluster, or impurity cluster, and approximates all other clusters by a non-interacting bath coupling to the impurity cluster, i.e., maps the original problem onto an impurity problem. The bath is determined self-consistently. The CDMFT neglects nonlocal self-energies between different clusters, i.e. $\Sigma_{\mu\nu} = 0$ for $\mu\neq\nu$ where $\mu$ and $\nu$ enumerate clusters, and further identifies the impurity cluster self-energy $\Sigma^I$ in the impurity problem with the local self-energy $\Sigma_{\mu\mu}$ in the original problem, i.e., $\Sigma_{\mu\mu} = \Sigma^I$. It is this assumption that makes the CDMFT break translation symmetry.

To establish a complete correspondence between the lattice problem and the impurity problem, the CDMFT introduces a Weiss field $\mathcal{G}$ defined by
\begin{equation}\label{eq:weiss}
\mathcal{G}^{-1} = (G_{\mu\mu})^{-1} + \Sigma_{\mu\mu},
\end{equation}
where $\Sigma_{\mu\mu}$ and $G_{\mu\mu}$ are actually independent of $\mu$ because of the translation symmetry of the superlattice. And $G_{\mu\mu}$ can be derived from
\begin{equation}\label{eq:lattG}
G({\bf{K}}, i\omega_n) = [i\omega_n - H_{0}({\bf{K}}) - \Sigma({\bf{K}}, i\omega_n)]^{-1}
\end{equation}
with ${\bf{K}}$ the superlattice momentum, $H_{0}({\bf{K}})$ the hopping coefficients of the model on the superlattice in the ${\bf{K}}$-space, and $\Sigma({\bf{K}}) = \Sigma^I$ independent of ${\bf{K}}$ by a simple derivation. Finally, with the Dyson equation of the impurity model $\Sigma^I = (G_0^I)^{-1} - (G^I)^{-1}$, the self-consistent equations of the CDMFT are closed by identifying the non-interacting Green's function of the impurity model $G_0^I$ with the Weiss field $\mathcal{G}$, i.e.,
\begin{equation}\label{eq:impGF0}
(G_0^I)^{-1} = \mathcal{G}^{-1}.
\end{equation}

In numerical realization of the CDMFT, the self-consistent equations are solved by an iterative algorithm. One begins with a guessed $\Sigma$, for example, the one obtained from other calculations or simply by setting $\Sigma = 0$. Substituting it into Eq.~(\ref{eq:lattG}) and Eq.~(\ref{eq:weiss}), one can obtain $G$ and $\mathcal{G}$, successively, and then $G_0^I$ from Eq.~(\ref{eq:impGF0}). Meanwhile, the impurity problem is independently solved by using an impurity solver such as ED or QMC to yield the Green's function of the impurity sites $G^I$. Thus, according to the Dyson equation of the impurity model, one obtains $\Sigma^I$, which then gives a new $\Sigma$ for the original lattice model. This process will be iterated until convergence.

When the ED is used as an impurity solver, the bath is discretized, i.e., described by a set of non-interacting sites coupling to the impurity sites. In a general form, the impurity model reads
\begin{equation}\label{impH}
\begin{aligned}
\hat{H}^I &= \sum_{i,j\in\text{cluster},\sigma}t_{ij}^C c_{i\sigma}^{\dagger}c_{j\sigma} + U\sum_{i\in\text{cluster}}n_{i\uparrow}n_{i\downarrow} \\
&+ \sum_{k\in\text{bath}}\varepsilon_{k}c_{k}^{\dagger}c_{k} + \sum_{i\in\text{cluster},k\in\text{bath}}(V_{ik}c_{i}^{\dagger}c_{k} + {\rm H.c}),
\end{aligned}
\end{equation}
where $t_{ij}^C$ and $U$ coincide with those in the original lattice model. And $\varepsilon_{k}$ and $V_{ik}$ enter into $G_0^I$ with
\begin{equation}\label{eq:fitG0I}
(G_0^I)^{-1}(i\omega_n) = i\omega_n - t^C - V(i\omega_n - \varepsilon)^{-1}V^{\dagger},
\end{equation}
where $\varepsilon$ is a diagonal matrix with diagonal elements $\varepsilon_k$, while the elements of matrix $V$ and $t^C$ are $V_{ik}$ and $t_{ij}^C$, respectively. To determine $\varepsilon_{k}$ and $V_{ik}$, we fit $(G_0^I)^{-1}(i\omega_n)$ with $\mathcal{G}^{-1}(i\omega_n)$, i.e., minimize the distance between $(G_0^I)^{-1}$ and $\mathcal{G}^{-1}$ by adjusting unknown parameters $\varepsilon_{k}$ and $V_{ik}$.~\cite{Caffarel94,He12}

Now we turn to the PCDMFT. As introduced above, the CDMFT breaks translation symmetry. By inspecting the self-consistent equations, we see that the lattice quantities $\Sigma({\bf{K}})$ and $G({\bf{K}})$ violate the translation symmetry of the original lattice because of directly setting equation $\Sigma({\bf{K}}) = \Sigma^I$. To restore the translation symmetry, in the PCDMFT we construct the lattice self-energy $\Sigma$ through periodizing $\Sigma^I$ as follows,~\cite{Biroli04,Capone04}
\begin{equation}\label{eq:periodize}
\Sigma({\bf{k}}) = \frac{1}{N_c} \sum_{\alpha\beta} e^{-i{\bf{k}} \cdot {\bf{R}}_\alpha} \Sigma_{\alpha\beta}^I e^{i{\bf{k}} \cdot {\bf{R}}_\beta},
\end{equation}
where $\alpha$ ($\beta$) enumerates primitive unitcells in the cluster, ${\bf{R}}_\alpha$ (${\bf{R}}_{\beta}$) is the lattice vector for primitive unitcell $\alpha$ ($\beta$), ${\bf{k}}$ is the lattice momentum, and $N_c$ denotes the number of primitive unitcells in the cluster. Accordingly, Eq.~(\ref{eq:lattG}) changes into
\begin{equation}\label{eq:PlattG}
G({\bf{k}}, i\omega_n) = [i\omega_n - H_{0}({\bf{k}}) - \Sigma({\bf{k}}, i\omega_n)]^{-1}.
\end{equation}
And in comparison with the Dyson equation of the impurity model, Eq.~(\ref{eq:weiss}) is changed into
\begin{equation}\label{eq:Pweiss}
\mathcal{G}^{-1} = (G_{\mu\mu})^{-1} + \Sigma^I,
\end{equation}
because $\Sigma_{\mu\mu} = \Sigma^I$ no longer holds after periodizing the self-energy. From now Eq. (\ref{eq:Pweiss}) will work with the Dyson equation of the impurity model to guarantee the iterative consistency.

Following the previous studies~\cite{Wu10,Liebsch11,Yu11,He12,Seki12,Liebsch13} in the literature, we choose a six-site ring as the impurity cluster for the honeycomb lattice, as shown in Fig.~\ref{fig:weiss}. However, this choice is troublesome for the periodization (Eq.~(\ref{eq:periodize})) since such an impurity cluster cannot be divided into a set of primitive unitcells. To overcome this difficulty, we first decompose Eq.~(\ref{eq:periodize}) formally into
\begin{eqnarray}
\Sigma_\gamma^P &\equiv& \frac{1}{N_c} \sum_{{\bf{R}}_\beta-{\bf{R}}_\alpha={\bf{R}}_\gamma} \Sigma_{\alpha\beta}^I ,  \label{eq:avg0}\\
\Sigma({\bf{k}}) &=& \sum_{\gamma} \Sigma_{\gamma}^P e^{i{\bf{k}} \cdot {\bf{R}}_\gamma}, \label{eq:avgper0}
\end{eqnarray}
and further generalize them into
\begin{eqnarray}
\Sigma_{\gamma;ab}^P &\equiv& \frac{1}{N_c} \sum_{{\bf{R}}_\beta-{\bf{R}}_\alpha={\bf{R}}_\gamma} \Sigma_{\alpha\beta;ab}^I  \label{eq:avg}, \\
\Sigma_{ab}({\bf{k}}) &=& \sum_{\gamma} \Sigma_{\gamma;ab}^P e^{i{\bf{k}} \cdot {\bf{R}}_{\gamma}}, \label{eq:avgper}
\end{eqnarray}
where $a$ and $b$ enumerate sites in primitive unitcells; especially, in $\Sigma_{\alpha\beta;ab}^I$ $a$ and $b$ denote the sites in unitcells $\alpha$ and $\beta$, respectively. When the impurity cluster can be divided into a set of primitive unitcells, Eqs.~(\ref{eq:avg}) and ~(\ref{eq:avgper}) are equivalent to Eqs.~(\ref{eq:avg0}) and ~(\ref{eq:avgper0}), respectively. However, when the impurity cluster not, the latter no longer works while the former still works. In this case, we still divide the impurity cluster into a set of primitive unitcells, in which some unitcells are partly included. We then assign zero to the impurity self energy with a site beyond the impurity cluster, represented by  Eq. (\ref{eq:avg}).

\section{Results}

We have applied the PCDMFT to the half-filled standard Hubbard model on the honeycomb lattice. To study the ground state properties, we used the ED as an impurity solver. Thus, we calculated the same physical quantities as those by the CDMFT reported in Ref.~\onlinecite{He12}, and then compared them between the PCDMFT and the CDMFT.  Specifically, the calculated quantities are the density of states, single-particle gap, staggered magnetization, and double occupancy, respectively shown in the following Figs. 2, 3, 4, and 5.

\begin{figure}[h]
  \includegraphics[width=\columnwidth]{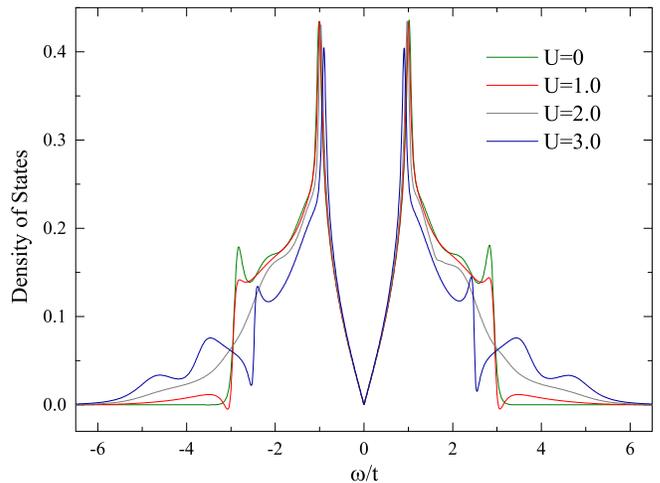}
  \caption{\label{fig:DOS} (Color online) Density of states (DOS) calculated by the PCDMFT. The linear behavior near the Fermi  energy shows a correlated Dirac semimetal.}
\end{figure}

\begin{figure}[h]
  \includegraphics[width=\columnwidth]{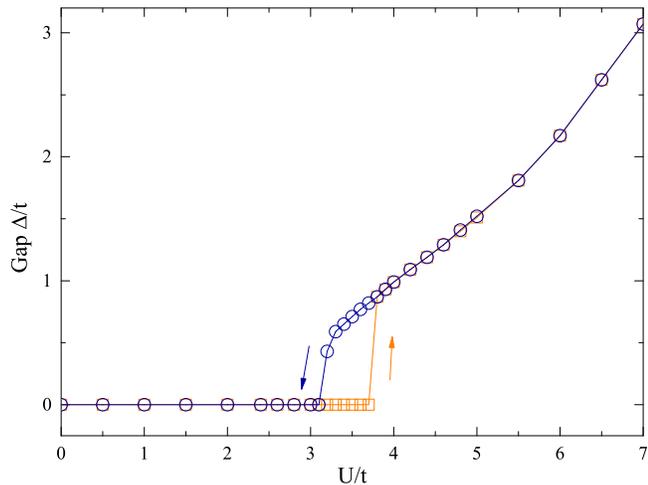}
  \caption{\label{fig:gap} (Color online) Single-particle gap $\Delta/t$ calculated by the PCDMFT. It shows hysteresis behavior around $U_{c} = 3.2 - 3.7t$, indicating a first-order quantum phase transition. For $U > U_c$, the system opens up a gap, indicating an insulating phase.}
\end{figure}

\begin{figure}[h]
  \includegraphics[width=\columnwidth]{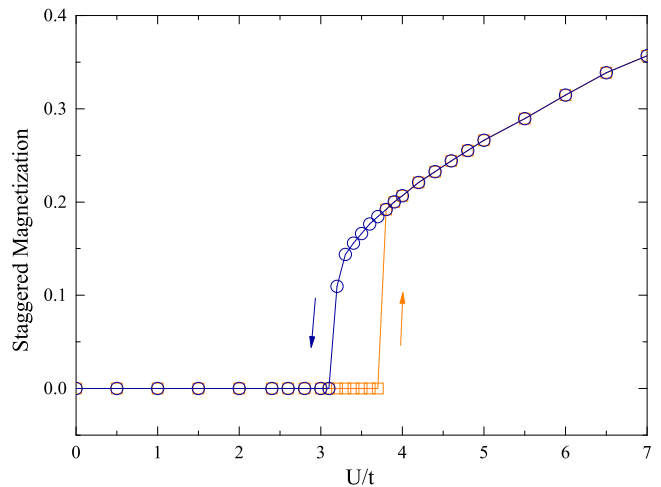}
  \caption{\label{fig:StaMag} (Color online) Staggered magnetization calculated by the PCDMFT. Hysteresis behavior is shown around $U_{c} = 3.2 - 3.7t$, indicating a first-order quantum phase transition. For $U > U_{c}$, the staggered magnetization is nonzero, indicating an antiferromagnetic phase.}
\end{figure}

For small interactions $U$, the density of states (Fig.~\ref{fig:DOS}) shows a perfect linear behavior near the Fermi energy, which means that the expected correlated Dirac semimetal is successfully produced. Correspondingly, both the single-particle gap $\Delta/t$ (Fig.~\ref{fig:gap}) and the staggered magnetization $m$ (Fig.~\ref{fig:StaMag}) are zero. In comparison, the CDMFT results in Ref.~\onlinecite{He12} show a nonzero $\Delta/t$ for any finite interaction.

\begin{figure}[h]
  \includegraphics[width=\columnwidth]{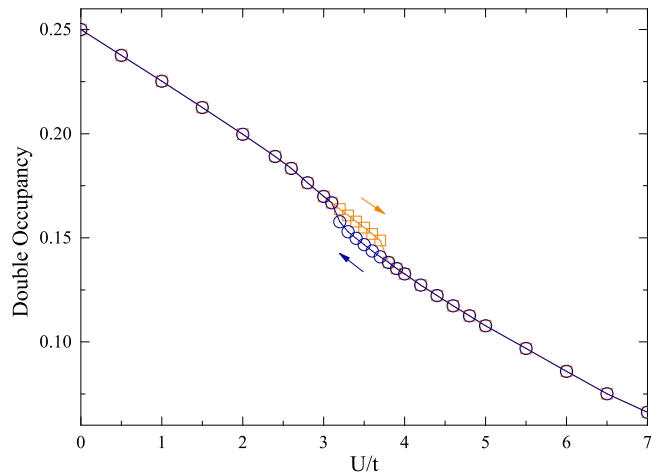}
  \caption{\label{fig:DouOcc} (Color online) Double occupancy calculated by the PCDMFT}
\end{figure}

As the interaction $U$ increases, the system encounters a first-order phase transition from the semimetallic phase to an antiferromagnetic insulating phase. The two phases coexist in the transition region of $U_{c1} < U < U_{c2}$ with $U_{c1} = 3.2t$ and $U_{c2} = 3.7t$. And the characteristic hysteresis behavior for a first-order phase transition shows up in the staggered magnetization (Fig.~\ref{fig:StaMag}), the single-particle gap $\Delta/t$ (Fig.~\ref{fig:gap}), and the double occupancy $D$ (Fig.~\ref{fig:DouOcc}) as functions of $U$, respectively. The converged result depends on the initial condition (the lattice self-energy $\Sigma$ in the present study) for the self-consistent iterations. Specifically, the converged result is the semimetallic phase if the initial $\Sigma$ is equal to the converged $\Sigma$ from a previous calculation with a small interaction or is simply set to zero; in contrast, the converged result is the antiferromagnetic insulating phase if the initial $\Sigma$ is equal to the converged $\Sigma$ from a previous calculation with a large interaction, as shown by the arrows in the figures.

For $U > U_{c2}$, the converged result no longer depends on the initial $\Sigma$ and is always the antiferromagnetic insulating phase, indicated by both the nonzero staggered magnetization (Fig.~\ref{fig:StaMag}) and the nonzero single-particle gap $\Delta/t$ (Fig.~\ref{fig:gap}). This is consistent with the result by the CDMFT in Ref.~\onlinecite{He12}.

As seen from the above, the main difference between the PCDMFT and CDMFT results~\cite{He12} is that the PCDMFT produces a correlated Dirac semimetal at small $U$, while the CDMFT produces an insulator for any finite $U$. The other results are similar, such as the double occupancy $D$ for all interactions. Nevertheless, $U_c$ is about 3.7t for the PCDMFT, which is remarkably closer to the values of $U_c = 3.8t$~\cite{Sorella12} and $U_c = 3.7t$~\cite{Assaad13} obtained from the large-scale QMC simulations, compared with that $U_c \approx 4.6t$ for the CDMFT~\cite{He12} and $U_c \approx 3.3t$ for the DCA~\cite{Liebsch13}.

\section{Discussion and conclusion}

The cluster dynamical mean-field theory (CDMFT) breaks the lattice translation symmetry by construction, which leads to a spurious phase that an insulating phase emerges at any small but finite interaction for the half-filled Hubbard model on the honeycomb lattice. To solve this issue, we have employed the periodized CDMFT (PCDMFT) which improves the CDMFT to partly restore the translation symmetry by periodizing the impurity cluster self-energy. It turns out that the PCDMFT successfully clears up the spurious insulating phase at small interaction $U$ and gives rise to a correlated Dirac semimetal instead. As $U$ increases, the system takes a first-order phase transition with $U_c \sim 3.7t$ from the semimetal to an antiferromagnetic insulator. And there is no a quantum spin liquid emerging near the transition. Basically, these results are in good agreement with the large-scale quantum Monte Carlo (QMC) results by Sorella et al.~\cite{Sorella12}

The PCDMFT inherits the CDMFT but meanwhile keeps translation symmetry like the dynamic cluster approximation (DCA). Thus it may be considered as a method interpolating between the CDMFT and the DCA. For the present model study, the DCA~\cite{Liebsch13} underestimates the transition point $U_c$ while the CDMFT~\cite{He12} overestimates it. In comparison, the PCDMFT gives the one most close to the large-scale QMC result~\cite{Sorella12,Assaad13}. This indicates that the PCDMFT makes a better balance between the descriptions of weak and strong correlation effects on the honeycomb lattice than the other two. Although these quantum cluster methods all guarantee to predict the same result by using an enough large impurity cluster, they may differ with each other significantly in prediction by using a small impurity cluster and thus should be used complementarily in practical studies.

\begin{acknowledgments}

This work is supported by National Natural Science Foundation of China (Grant Nos. 11474356 and 11190024), National Program for Basic Research of MOST of China (Grant No. 2011CBA00112), and Fundamental Research Funds for the Central Universities and the Research Funds of Renmin University of China (15XNH067). Computational resources were provided by the Physical Laboratory of High Performance Computing in RUC.

\end{acknowledgments}

\bibliography{mybib}

\end{document}